\documentclass[aps,preprint,showpacs,groupedaddress,floatfix]{revtex4}
\usepackage{graphicx}
\usepackage{dcolumn}
\usepackage{bm}

\begin{document}
\title{Relativistic QRPA description of isoscalar compression modes in 
open-shell nuclei in the A$\approx$60 mass region }
\author{N. Paar\footnote{Electronic address: nils.paar@physik.tu-darmstadt.de}}
\affiliation{
Institut f\" ur Kernphysik, Technische Universit\" at Darmstadt,
Schlossgartenstrasse 9,
D-64289 Darmstadt, Germany}
\affiliation{Physics Department, Faculty of Science, University of Zagreb,
Croatia}
\author{D. Vretenar}
\author{T. Nik\v si\' c}
\affiliation{Physics Department, Faculty of Science, University of Zagreb, 
Croatia}
\author{P. Ring}
\affiliation{Physik-Department der Technischen Universit\"at M\"unchen, 
D-85748 Garching,
Germany}
\date{\today}

\begin{abstract}
Very recent inelastic $\alpha$-scattering data on the 
isoscalar monopole and dipole strength distributions in 
$^{56}$Fe, $^{58}$Ni, and $^{60}$Ni are analyzed in the 
relativistic quasiparticle random-phase approximation 
(RQRPA) with the DD-ME2 effective nuclear interaction
(nuclear matter compression modulus K$_{nm}= 251$ MeV).
In all three nuclei the calculation nicely reproduces 
the observed asymmetric shapes of the monopole strength, 
and the bimodal structure of the dipole strength distributions.
The calculated centroid and mean energies are in very good
qualitative agreement with the experimental values both for the 
monopole, and for the low- and high-energy components of 
the dipole transition strengths. It is noted, however, 
that while DD-ME2 reproduces in detail the excitation 
energies of the giant monopole resonances (GMR) in nuclei 
with $A \ge 90$, the theoretical centroids are systematically 
above the experimental values in lighter nuclei with $A \leq 60$.
The latter can be reproduced with an effective interaction  
with a lower value of K$_{nm} \approx $ 230 MeV but,
because of the asymmetric shapes and pronounced fragmentation 
of the monopole strength distributions, isoscalar GMR data in 
light nuclei cannot provide accurate estimates of the nuclear 
matter compression modulus.
\end{abstract}

\maketitle

\bigskip \bigskip

Experimental excitation energies of compressional (monopole and
dipole) vibrational modes in atomic nuclei can in principle be used 
to deduce the value of the nuclear matter compression
modulus K$_{nm}$ \cite{Bla.80}. This quantity is related
to the curvature of the nuclear matter equation of state at 
the saturation point, and controls basic properties of atomic nuclei, 
the structure of neutron stars, the dynamics
of heavy-ion collisions and of supernovae explosions. The nuclear 
matter compressibility cannot be measured directly, but rather deduced 
from a comparison of experimental excitation energies of isoscalar giant 
monopole resonances (ISGMR), with the corresponding values predicted by 
microscopic nuclear effective interactions 
characterized by different values of K$_{nm}$.

Inelastic $\alpha$-scattering experiments have been used in 
high precision studies of the systematics of ISGMR in nuclei 
with $A \ge 90$. Nuclear structure models provide a consistent
description of the shapes of strength distributions and the
mass dependence of excitation energies, and thus relate the ISGMR  
to the nuclear compressibility and to the nuclear matter 
compression modulus. There is much less experimental information, 
and only few microscopic theoretical analyses of the 
structure of compressional modes in lighter nuclei with $A < 90$. 
While in heavy nuclei the shape of the ISGMR strength distribution is
typically symmetric, for $A < 90$ the ISGMR display asymmetric shapes 
with a slower slope on the high energy side of the peak, and with 
a further decrease of the mass number the ISGMR strength 
distributions become strongly fragmented. An interesting question, 
of course, is whether studies of compressional vibrations in 
lighter nuclei can provide additional information on the 
nuclear matter compression modulus. Namely, K$_{nm}$ corresponds to
bulk nuclear compressibility, whereas one expects that 
surface compressibility plays an increasingly important role in the
structure of ISGMR in lighter systems. In a very recent study \cite{Lui.06}
isoscalar giant resonances in $^{56}$Fe, $^{58}$Ni, and $^{60}$Ni 
have been studied with small-angle inelastic $\alpha$-scattering. 
In particular, most of the expected isoscalar E0 has been identified 
below 40 MeV excitation energy, and between 56\% and 72\% of the 
isoscalar E1 strength has been located in these nuclei. It was 
noted, however, that there are no specific microscopic calculations 
of E0 and E1 strength distributions in $^{56}$Fe and $^{60}$Ni. 
The mass dependence of the ISGMR excitation energies between 
$A = 40$ and $A = 90$ was thus compared with results of 
leptodermous expansions based on Hartree-Fock + RPA calculations with 
Skyrme interactions~\cite{Nay.90}, and constrained 
relativistic mean-field calculations \cite{Cho.97}. The purpose of this
work is to perform fully self-consistent relativistic quasiparticle
random-phase approximation (RQRPA) calculations of isoscalar E0 and 
E1 strength distributions in $^{56}$Fe, $^{58}$Ni, and $^{60}$Ni, 
using a modern effective density-dependent interaction which is known
to reproduce the systematics of compressional modes in heavier nuclei 
with $A \ge 90$.
 
Theoretical studies of nuclear compressional modes in the last decade 
have employed the fluid dynamics approach~\cite{Kol.00},
the Hartree-Fock plus random phase
approximation (RPA)~\cite{Ham.97,Shl.02,Agr.03,Col.04}, 
the RPA based on separable Hamiltonians~\cite{Kva.03}, linear response
within a stochastic one-body transport theory~\cite{Lac.01}, 
the relativistic transport approach~\cite{Yil.05},
and the self-consistent relativistic RPA~\cite{MGW.01,Pie.01,Pie.02,VNR.03}.
As has been pointed out by Shlomo et al., however, most current 
implementations of the non-relativistic RPA are
not self-consistent, and based on numerous approximations~\cite{SS.02}.
Very recent studies have emphasized the importance of a 
fully self-consistent description of ISGMR, and confirmed that
the low value of K$_{nm}= 210-220$ MeV, previously obtained with Skyrme
functionals, is an artefact of the inconsistent implementation of effective
interactions \cite{Col.04,CG.04}. The excitation energies of the ISGMR 
in heavy nuclei are thus best described with Skyrme and Gogny effective 
interactions with K$_{nm} \approx$ 235 MeV. In Ref.~\cite{Agr.03} it has 
been shown that it is also possible to construct Skyrme forces that fit 
nuclear ground state properties and reproduce ISGMR energies, but with 
K$_{nm}$ $\approx 255$ MeV. In Ref.~\cite{Col.04} a new set of Skyrme forces 
was constructed that spans a wider range of values of K$_{nm}$ and 
the symmetry energy at saturation density $a_4$.  RPA calculations with 
these forces have shown that the ISGMR data are best 
reproduced with K$_{nm} = 230-240$ MeV, whereas higher values of K$_{nm}$ 
would require unrealistically large value of $a_4$. On the other hand, 
it appears that in the relativistic framework the interval of allowed
values for K$_{nm}$ is more restricted. A recent relativistic RPA 
analysis based on modern effective Lagrangians with explicit 
density dependence of the meson-nucleon vertex functions, has shown that 
only effective interactions with K$_{nm}= 250-270$ MeV reproduce 
the experimental excitation energies of ISGMR in medium-heavy and 
heavy nuclei, and that K$_{nm} \approx 250$ MeV represents the lower 
limit for the nuclear matter compression modulus of relativistic
mean-field interactions \cite{VNR.03}. 

Data on the compressional isoscalar giant dipole resonance (ISGDR) could 
also be used to constrain the range of allowed values of 
K$_{nm}$ \cite{Tex.01,Tex.04}. The problem, however, is that the isoscalar
E1 strength distributions display a characteristic bimodal structure with 
two broad components: one in the low-energy region close to the isovector
giant dipole resonance (IVGDR) 
($\approx 2 \hbar \omega$), and the other at higher
energy close to the electric octupole resonance ($\approx 3 \hbar \omega$).
Theoretical analyses have shown that only the high-energy component 
represents compressional vibrations \cite{Colo.00,VWR.00}, whereas
the broad structure in the low-energy region corresponds to vortical 
nuclear flow associated with the toroidal dipole 
moment \cite{BMS.93,VPRN.02,Mis.06}. However, as has also been pointed
out in the recent study of the interplay between compressional and 
vortical nuclear currents \cite{Mis.06}, a strong mixing between 
compressional and vorticity vibrations in the isoscalar E1 states 
can be expected up to the highest excitation energies in the region 
$\approx 3 \hbar \omega$. Nevertheless, models which use effective 
interactions with K$_{nm}$ adjusted to ISGMR excitation energies 
in heavy nuclei, also reproduce the structure of the high-energy 
portion of ISGDR data~\cite{SS.02,Uch.03,Uch.04}.

In this work the isoscalar E0 and E1 strength distributions for the 
open-shell nuclei $^{56}$Fe, $^{58}$Ni, and $^{60}$Ni are calculated 
in the relativistic quasiparticle random-phase approximation (RQRPA), 
formulated in the canonical single-nucleon basis of the 
relativistic Hartree-Bogoliubov (RHB) model \cite{Paa.03}.
The RQRPA model is fully self-consistent: the same interactions, in
the particle-hole and particle-particle channels, are used both in
the RHB equations that determine the canonical quasiparticle basis,
and in the RQRPA equations. In both channels the same strength
parameters of the interactions are used in the RHB and RQRPA
calculations. This is an essential feature of the RHB+RQRPA approach
and it ensures that RQRPA amplitudes do not contain spurious
components associated with the mixing of the nucleon number in the
RHB ground state, or with the center-of-mass translational motion.

\begin{figure}
\includegraphics[scale=0.6]{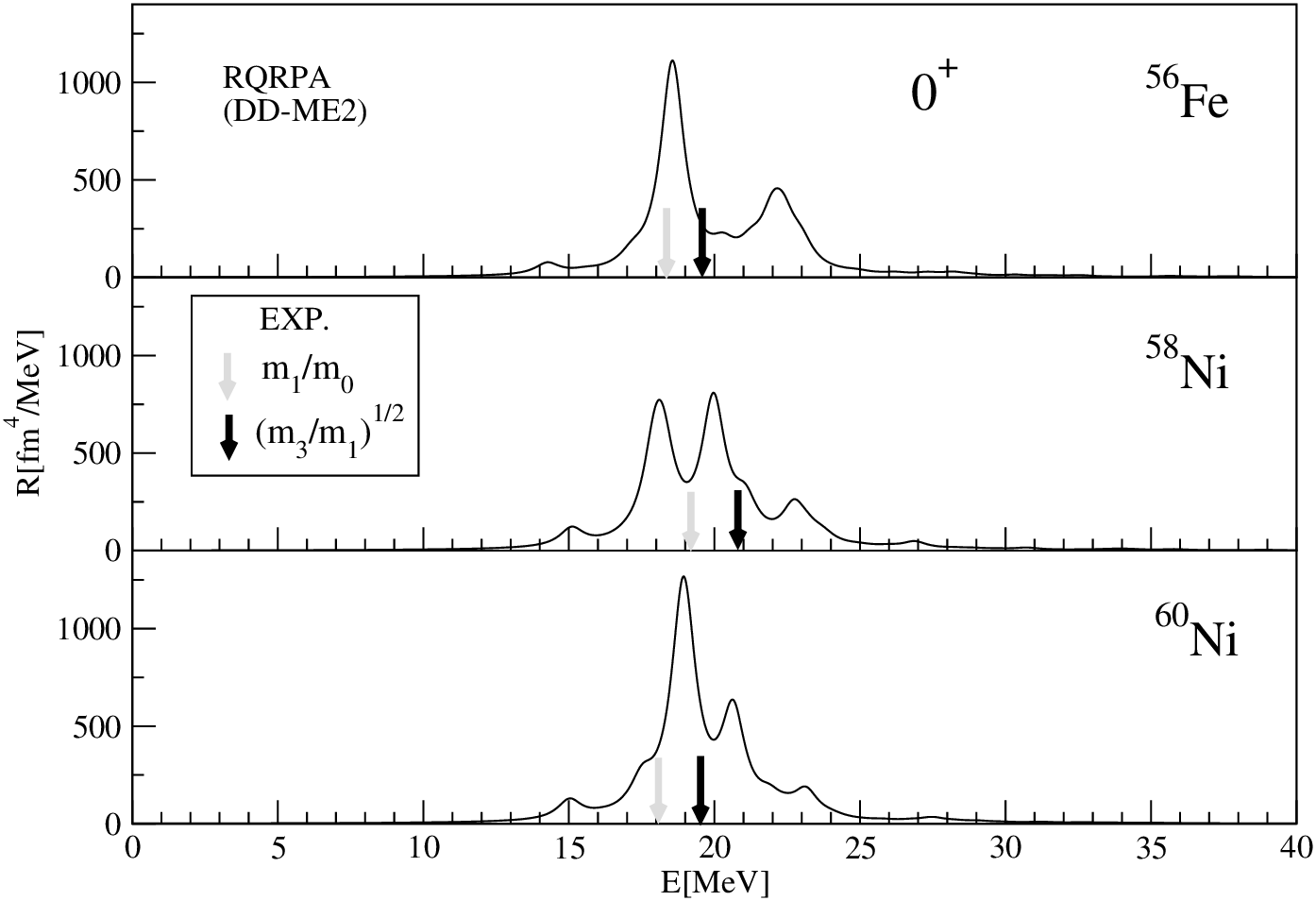}
\vspace{1cm}
\caption{The RHB+RQRPA isoscalar monopole strength distributions in
$^{56}$Fe, $^{58}$Ni, and $^{60}$Ni calculated with the
density-dependent effective interaction DD-ME2. 
The experimental centroid ($m_1/m_0$)
and mean ($\sqrt{m_3/m_1}$) energies obtained from $(\alpha,\alpha')$
scattering~\protect\cite{Lui.06} are denoted by grey
and black arrows, respectively.}
\label{Fig1}
\end{figure}

In Fig.~\ref{Fig1} we display the isoscalar monopole strength 
distributions for $^{56}$Fe, $^{58}$Ni, and $^{60}$Ni. The RHB+RQRPA 
calculation has been performed with the DD-ME2 effective
interaction \cite{Lal.05} in the particle-hole channel and, as in most 
applications of the RHB model \cite{Vre.05}, the finite-range 
Gogny force has been used in the particle-particle channel. DD-ME2 
belongs to a new class of relativistic effective nuclear interactions 
with density-dependent meson-nucleon vertex functions. In a number of 
recent studies it has been shown that this type of 
effective interactions provides a realistic description of asymmetric 
nuclear matter, neutron matter and finite spherical and 
deformed nuclei. These interactions allow for a 
softer equation of state of nuclear 
matter (i.e. lower incompressibility) and a lower value of the symmetry 
energy at saturation. In addition to nuclear matter and ground-state 
properties of spherical nuclei, the parameters of DD-ME2 have been adjusted 
to the excitation energies of the ISGMR and IVGDR in $^{208}$Pb. For  
DD-ME2 the nuclear matter compression modulus amounts 
K$_{nm}= 251$ MeV. The strength distributions in 
Fig.~\ref{Fig1} can be compared with the data from Ref.~\cite{Lui.06} 
(Figs. 8, 9 and 10). In all three nuclei the calculation predicts 
asymmetric shapes for the isoscalar E0 strength distributions, in 
agreement with data. In particular, an additional tail in the
transition strength is obtained above the main ISGMR peaks
for $^{56}$Fe and $^{60}$Ni. For $^{58}$Ni most of the 
strength is distributed over two major peaks, with an additional
pronounced high-energy tail. The arrows denote the positions of 
the experimental centroid ($\bar{E}_1=m_1/m_0$) and 
mean energies ($\bar{E}_3=\sqrt{m_3/m_1}$), where $m_k=\int E^kR(E)dE$ are the 
energy moments, and $R(E)$ is the transition strength distribution function.
We note that in all three nuclei the main ISGMR peak predicted by the RQRPA 
calculation is located in the narrow energy window between the  
$\bar{E}_1$ and $\bar{E}_3$ experimental energies.

\begin{figure}
\includegraphics[scale=0.6]{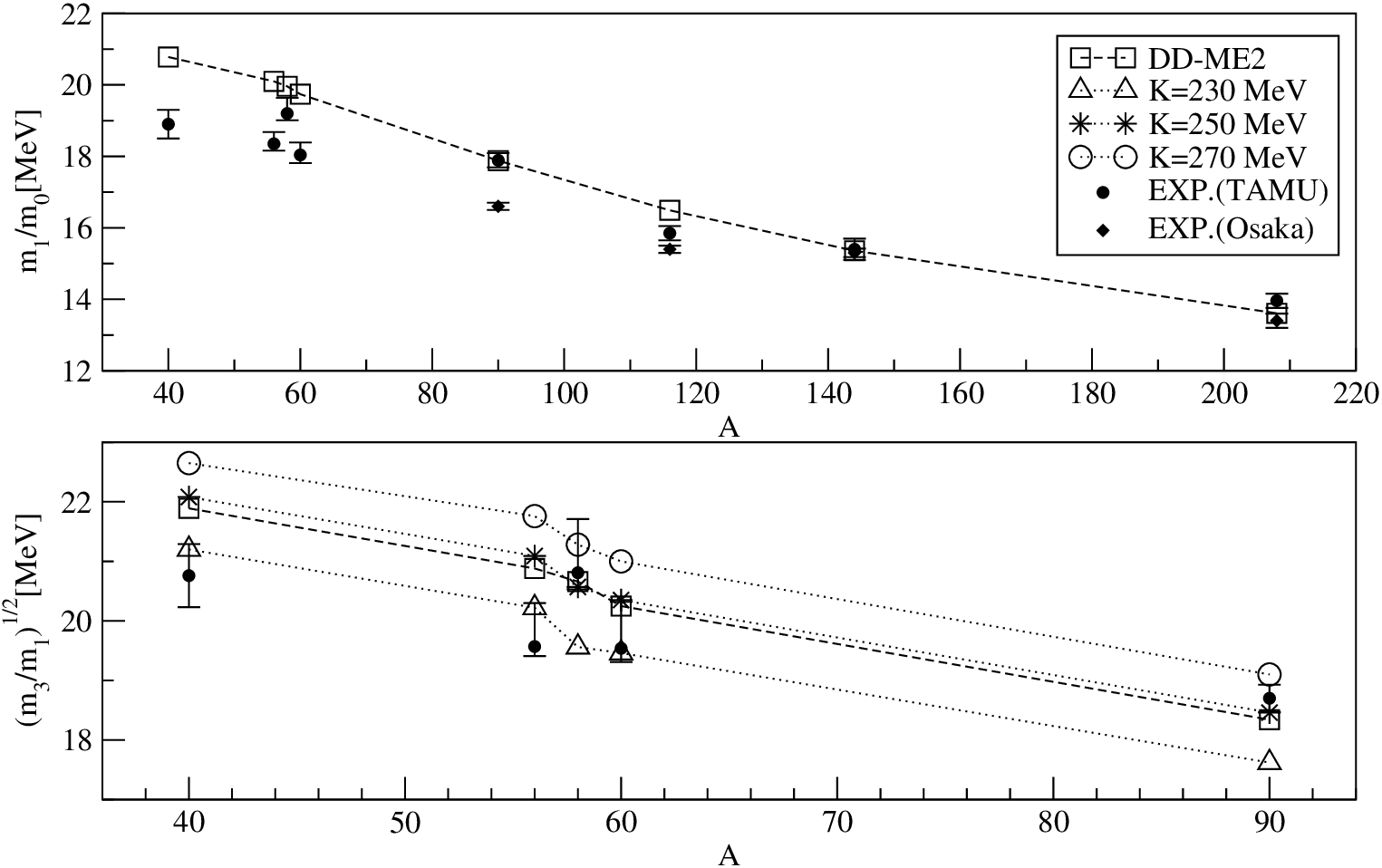}
\vspace{1cm}
\caption{In the mass region $40\leq A \leq 208$ the RQRPA results for
the ISGMR centroid energies $(m_1/m_0)$, calculated with the 
relativistic DD-ME2 effective interaction, are plotted as a 
function of mass number and compared with data from the
Texas A\&M University (TAMU)~\protect\cite{You.97,You.99,You.04,Lui.06} and
Osaka~\protect\cite{Uch.03,Uch.04} compilations (upper panel).
In the lower panel the calculated ISGMR excitation energies $\sqrt{m_3/m_1}$ 
for several medium-mass nuclei are shown in comparison with the data 
from Ref.~\cite{Lui.06}. In addition to the DD-ME2 interaction, 
three additional effective interactions with the values of the 
nuclear matter compressibility K$_{nm}$=230, 250, and 270 MeV  
\protect\cite{VNR.03} have been used in the RHB+RQRPA 
calculation of the ISGMR strength distributions.
}
\label{Fig2}
\end{figure}

In the upper panel of Fig.~\ref{Fig2} we plot the RHB+RQRPA results for
the ISGMR centroid energies $(m_1/m_0)$ of a series of spherical nuclei 
from $^{40}$Ca to $^{208}$Pb, calculated with the DD-ME2 effective 
interaction, in comparison with data from the 
Texas A\&M University (TAMU)~\protect\cite{You.97,You.99,You.04,Lui.06} 
and Osaka~\protect\cite{Uch.03,Uch.04} compilations. We note that the 
latter data correspond to peak energies and, especially in nuclei in which 
a high-energy tail is found above the main peak, these values should be
somewhat below the TAMU centroid energies. The agreement between the 
excitation energies calculated with DD-ME2 and the TAMU data is 
remarkable for nuclei with $A \ge 90$, whereas the theoretical 
centroids are systematically above the experimental values in lighter 
nuclei. The origin of this discrepancy is not understood, but it could 
be due to the fact that in light nuclei the surface incompressibility 
plays a more important role in determining the ISGMR, 
whereas K$_{nm}$ represents the volume incompressibility.
The former quantity is seldom taken into account when adjusting the 
parameters of an effective interaction and, therefore, we do not really
expect that DD-ME2 can reproduce in detail the moments of asymmetric and 
even fragmented isoscalar E0 strength distributions in light nuclei with 
$A \leq 60$. 

It seems that data on ISGMR in light nuclei are not very useful in 
extracting information on the nuclear matter compression modulus K$_{nm}$.
Nevertheless, we have tried to reproduce these data with few additional
effective interactions. In the recent analysis of nuclear matter 
incompressibility in the relativistic mean-field framework \cite{VNR.03},
families of density-dependent interactions with different values 
of the nuclear matter compression modulus K$_{nm}$ and symmetry energy 
at saturation (volume asymmetry) a$_4$, were adjusted to reproduce 
nuclear matter and ground-state properties of spherical nuclei. 
By performing fully consistent RRPA/RQRPA calculations of isoscalar E0 
and isovector E1 strength distributions in spherical nuclei with 
$A \ge 90$, it has been shown that the comparison with data 
restricts the values of K$_{nm}$ to $\approx 250 - 270$ MeV, 
and the range of volume asymmetry to 32 MeV $\leq a_4 \leq$ 36 MeV. 
A weak correlation between a$_4$ and K$_{nm}$ was found, i.e. 
interactions with lower volume asymmetry allow for slightly lower 
values of K$_{nm}$. Therefore in addition to DD-ME2, the 
family of interactions with a$_4$ = 32 MeV and  K$_{nm}$=230, 250, 
and 270 MeV~\cite{VNR.03} has been used in a RHB+RQRPA calculation
of ISGMR in $^{40}$Ca,
$^{56}$Fe, $^{58}$Ni, $^{60}$Ni, and $^{90}$Zr. The resulting mean
energies $\bar{E}_3=\sqrt{m_3/m_1}$ are plotted in the lower panel 
of Fig.~\ref{Fig2}, in comparison with data from 
Refs.~\cite{You.97,You.99,Lui.06}. We notice that while DD-ME2 and 
the K$_{nm}$=250 MeV effective interaction reproduce the 
experimental value $\bar{E}_3$ for $^{90}$Zr, data in lighter nuclei 
are better described by the effective interaction with 
K$_{nm}$=230 MeV, except possibly for $^{58}$Ni, but for this nucleus 
the experimental $\bar{E}_3$ differs considerably from the 
values in the neighboring $^{56}$Fe and $^{60}$Ni \cite{Lui.06}. 

\begin{figure}
\includegraphics[scale=0.6]{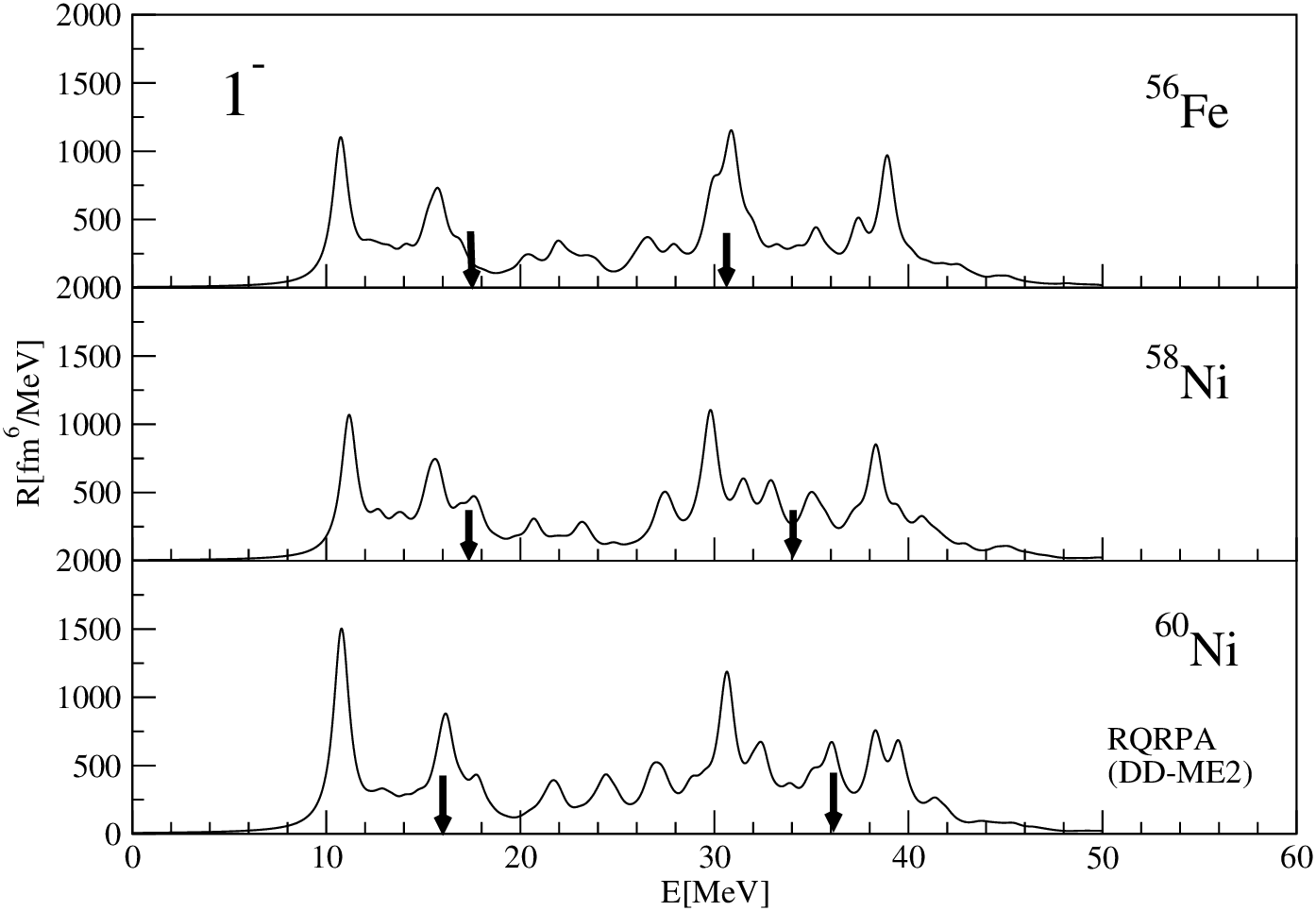}
\vspace{1cm}
\caption{The RHB+RQRPA isoscalar dipole transition strength in
$^{56}$Fe, $^{58}$Ni, and $^{60}$Ni calculated with
DD-ME2 effective interaction. The arrows denote the positions 
of the experimental centroid energies of the low- and high-energy 
components \protect\cite{Lui.06}.}
\label{Fig3}
\end{figure}

For the DD-ME2 effective interaction, the RHB+RQRPA isoscalar dipole 
transition strength distributions in $^{56}$Fe, $^{58}$Ni, and $^{60}$Ni 
are shown in Fig.~\ref{Fig3}. In all three nuclei the E1 strength 
is strongly fragmented and distrubuted over a wide range of 
excitation energy between 10 MeV and 40 MeV, in agreement 
with the experimental results of Ref.~\cite{Lui.06}. In the experiment 
between 56\% and 72\% of the isoscalar E1 strength has been located in
these nuclei below 40 MeV excitation energy, and some missing strength 
probably lies at higher energies. Similarly to the results obtained for 
heavier nuclei \cite{Colo.00,VWR.00,VPRN.02}, the E1 strength is 
basically concentrated in two broad structures: one in the region 
10 MeV $\leq$ E$_x$ $\leq$ 20 MeV, and the high-energy component 
above 25 MeV and extending above 40 MeV excitation energy. Only the 
high-energy portion of the calculated E1 strength is sensitive to the 
the nuclear matter compression modulus of the effective interaction. 
In a number of recent theoretical studies \cite{BMS.93,VPRN.02,Mis.06} 
it has been shown that the low-lying E1 strength mostly corresponds 
to vortical flow (dipole toroidal mode), although a strong mixture 
of compressional and vortical velocity fields is predicted in the 
intermediate and high-energy region. 

\begin{figure} 
\includegraphics[scale=0.6]{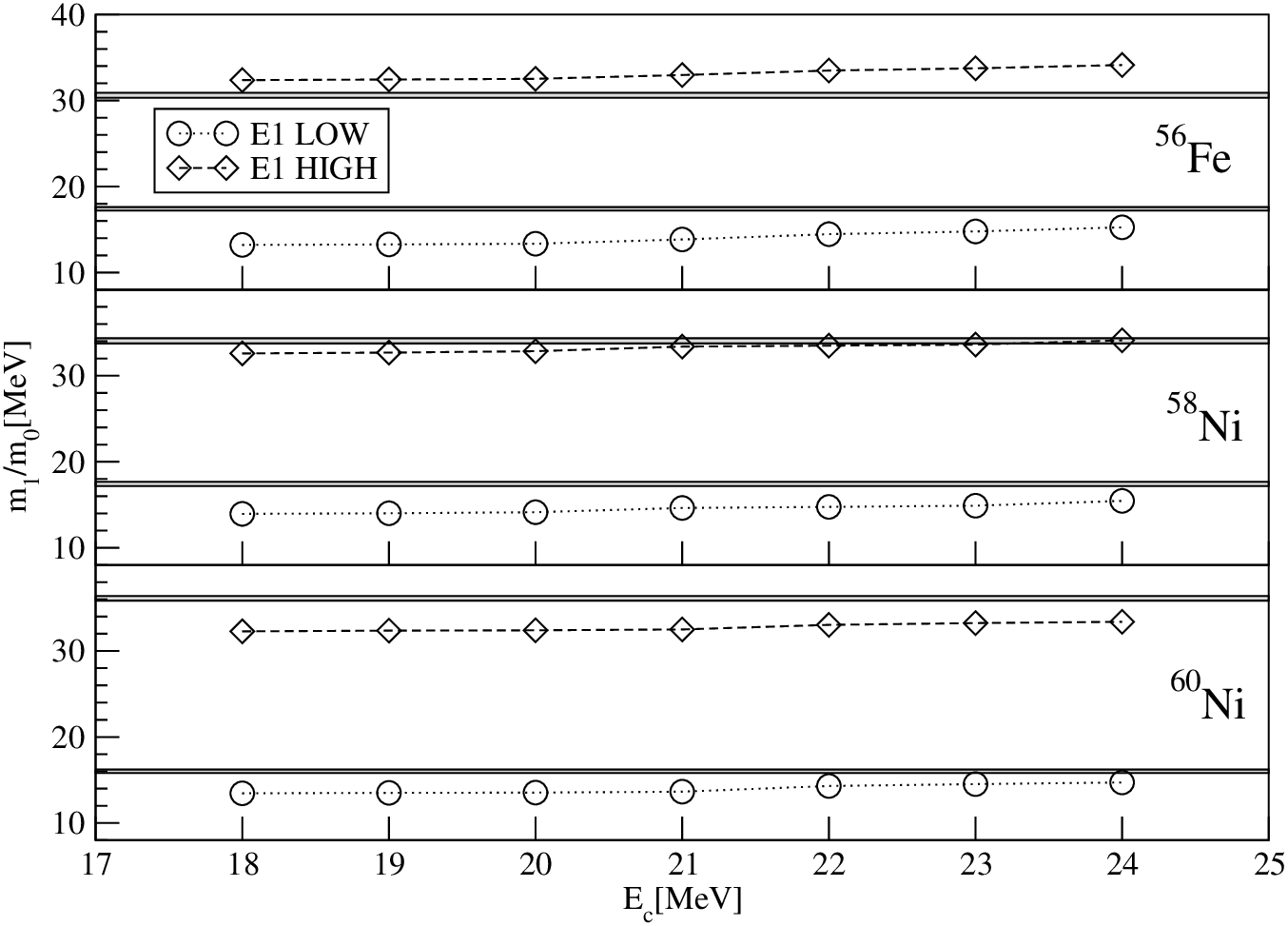}
\vspace{1cm}
\caption{Comparison of the RHB+RQRPA (circles and diamonds)
and experimental (thick lines) \protect\cite{Lui.06}
centroid energies of the low- and high-energy components 
of the isoscalar dipole transition strength in 
$^{56}$Fe, $^{58}$Ni, and $^{60}$Ni. The results 
are plotted for different values of E$_c$, the parameter which 
separates the low- and high-energy regions.}
\label{Fig4}
\end{figure}

In Fig.~\ref{Fig3} the thick arrows denote the locations of the 
experimental centroid energies ($m_1/m_0$) in the 
low- and high-energy regions of the isoscalar E1 strength in 
$^{56}$Fe, $^{58}$Ni, and $^{60}$Ni \cite{Lui.06}.
These are compared in Fig.~\ref{Fig4} with the theoretical 
values of the centroids of the low- and high-energy components, 
for different values of E$_c$, the somewhat arbitrary parameter 
which separates the low- and high-energy regions. We notice a 
good qualitative agreement between the calculated and experimental 
centroids in the high-energy region, especially taking into 
account that the E1 strength above E$_x = 40$ MeV has not been 
observed in the experiment. In the low-energy region, however,  
the theoretical centroid energies are systematically below 
the experimental values by $\approx$1--4 MeV, depending  
on the choice of E$_c$. This effect is in agreement 
with previous RRPA calculations in heavier nuclei \cite{VPRN.02}, 
and supports the picture of pronounced mixing between 
compressional and vorticity vibrations in the intermediate 
region of excitation energies.  

In conclusion, we have performed RHB+RQRPA calculations 
of the isoscalar monopole and dipole strength distributions 
in A$\approx$60 nuclei. The results obtained with the DD-ME2 
effective interaction (nuclear matter compression modulus 
K$_{nm}= 251$ MeV) have been compared with very recent data on 
the E0 and E1 strength distribution in 
$^{56}$Fe, $^{58}$Ni, and $^{60}$Ni \cite{Lui.06}. For the 
isoscalar monopole resonance we find very good qualitative agreement 
between theory and experiment, both for the asymmetric 
shapes of the strength distributions, as well as for 
centroid ($\bar{E}_1=m_1/m_0$) and mean energies 
($\bar{E}_3=\sqrt{m_3/m_1}$). It has been noted, however, 
that while there is an excellent agreement between the ISGMR
excitation energies calculated with DD-ME2 and the data for nuclei 
with $A \ge 90$, the theoretical centroids are systematically above
the experimental values in lighter nuclei with $A \leq 60$. Even 
though because of asymmetric shapes and pronounced fragmentation, 
ISGMR data in light nuclei are probably not very useful for 
extracting information on the nuclear matter compression modulus, 
we have shown that the ISGMR centroids in nuclei with $A \leq 60$ 
are better described with an effective interaction similar to 
DD-ME2, but with a lower value of K$_{nm} \approx $ 230 MeV.
The isoscalar E1 strength distributions calculated with DD-ME2 
are in good agreement with the experimental results \cite{Lui.06}, 
and reproduce the observed bimodal structure with two broad 
components in the $2 \hbar \omega$ and $3 \hbar \omega$ 
energy regions. The calculated centroid energies of the 
low- and high-energy E1 components in $^{56}$Fe, $^{58}$Ni, 
and $^{60}$Ni qualitatively reproduce the experimental values 
obtained from small-angle inelastic $\alpha$-scattering data.
\bigskip

\leftline{\bf ACKNOWLEDGMENTS}
\noindent
This work has been supported in part by the Bundesministerium
f\"ur Bildung und Forschung under project 06 TM 193, by the
Gesellschaft f\" ur Schwerionenforschung (GSI) Darmstadt, and
by the Alexander von Humboldt Stiftung. 
\bigskip
\newpage
\vspace{1.0cm}
\vspace{1.0cm}
\vspace{1.0cm}
\end{document}